\documentclass[article, twocolumn,
   aps, pra,
  amsmath,amssymb,
  longbibliography,
  ]{revtex4-1}
\usepackage{graphicx,color}
\usepackage{amsmath}
\usepackage{natbib}
\usepackage{epsfig}
\begin{document}

\title{Generalized Rosenfeld-Tarazona scaling and high-density specific heat of simple liquids}

\author{S. A. Khrapak}\email{Sergey.Khrapak@gmx.de}
\affiliation{Joint Institute for High Temperatures, Russian Academy of Sciences, 125412 Moscow, Russia}
\author{A. G. Khrapak}
\affiliation{Joint Institute for High Temperatures, Russian Academy of Sciences, 125412 Moscow, Russia}

\begin{abstract}
The original Rosenfeld-Tarazona (RT) scaling of the excess energy in simple dense fluids predicts a $\propto T^{3/5}$ thermal correction to the fluid Madelung energy. This implies that the excess isochoric heat capacity scales as $C_{\rm v}^{\rm ex}\propto T^{-2/5}$. Careful examination performed in this paper demonstrates that the exponent $-2/5$ is not always optimal. For instance, in the Lennard-Jones fluid in some vicinity of the triple point, the exponent $-1/3$ turns out to be more appropriate. The analysis of the specific heat data in neon, argon, krypton, xenon, and liquid mercury reveals that no single value of the exponent exists, describing all the data simultaneously. Therefore we propose a generalized RT scaling in the form  $C_{\rm v}^{\rm ex}\propto T^{-\alpha}$, where $\alpha$ is a density- and material-dependent adjustable parameter. The question concerning which material properties and parameters affect the exponent $\alpha$ and whether it can be predicted from general physical arguments requires further investigation.
\end{abstract}

\date{\today}

\maketitle

\section{Introduction}

Thermodynamic and transport properties of the liquid state have received renewed attention in recent years. Among the important current research topics are collective excitations and their relations to liquid thermodynamics~\cite{BolmatovSciRep2012,YangPRL2017,BrykJCP2017,
KhrapakSciRep2017,KryuchkovSciRep2019,ProctorPoF2020,KryuchkovPRL2020,
BaggioliPRE2021,SchirmacherPRE2022,BolmatovJPCL2022,TrachenkoBook}, gas-like to liquid-like dynamical crossover in supercritical fluids (Frenkel line in the phase diagram)~\cite{GorelliPRL2006,Simeoni2010,
BrazhkinPRE2012,BrazhkinUFN2012,BrazhkinPRL2013,
GorelliSciRep2013,BrykJPCL2017,BrazhkinJPCB2018,
BrykJPCB2018,BellJCP2020,ProctorJPCL2019,BellJPCL2021,
CockrellPR2021,KhrapakJCP2022,RanieriNatCom2024}, vibrational paradigm of liquid transport properties~\cite{KhrapakPhysRep2024,KhrapakPRE01_2021,KhrapakPoP08_2021,KhrapakMolecules12_2021,
KhrapakPPR2023,KhrapakPRE2023}, as well as some other properties of liquid transport~\cite{OhtoriChemLett2020,KhrapakPRE04_2021,KhrapakPRE10_2021,KhrapakJPCL2022,
Ranieri2021,KhrapakJMolLiq2022,LiJCP2021,BaranJCP2023,KhrapakJCP2023,
TrachenkoPRB2021,KhrapakPoF2022,KhrapakPoF2023,KhrapakJCP2024}. Since no general theory of liquid transport properties exists and very unlikely to be ever developed, various approximate relations and scaling relationships turn out to be helpful in estimating and predicting transport coefficients. One such useful scaling is the excess entropy scaling of macroscopically reduced transport coefficients proposed by Rosenfeld~\cite{RosenfeldPRA1977}. According to this scaling the reduced self-diffusion,viscosity and thermal conductivity coefficients of simple fluids can be expressed as exponential functions of the reduced excess entropy (entropy per atom in units of $k_{\rm B}$), $s_{\rm ex}=s-s_{\rm id}$, where $s_{\rm id}$ is the reduced entropy of an ideal gas at the same temperature and density. Many simple and not so simple systems conform to the approximate excess entropy scaling (e.g. binary mixtures, ionic substances, small molecules, polymers, and active-matter systems; see for instance Refs.~\cite{IngebrigtsenJCP2013,BellNatComm2020,GhaffarizadehJPCL2022} and references therein). There are also counterexamples, where the original excess entropy scaling is not applicable (among known exceptions are water models, the Gaussian core model, the Hertzian model, the soft repulsive-shoulder-potential model, models with flexible molecules, etc. ~\cite{KrekelbergPRE12_2009,FominPRE2010,DyreJPCB2014}). An excellent overview of excess entropy scaling and related topics is given in Ref.~\cite{DyreJCP2018}.   

To utilize the predictive power of the excess entropy scaling (in cases it is applicable), access to the thermodynamic properties is required. A simple practical approach was proposed by Rosenfeld and Tarazona (RT), who derived a unified analytical description of classical bulk solids and fluids, predicting correctly major features of their equations of state and freezing
conditions~~\cite{RosenfeldMolPhys1998}. Starting with a fundamental-measure hard-sphere reference functional they were able to demonstrate that the high density expansions for the potential energy are dominated by a
static-lattice Madelung term and the harmonic $1.5 T$ correction for solids and fluid Madelung energy with a $\propto T^{3/5}$ thermal correction for fluids. Moreover, both expansions originate from the same singularity in the
hard sphere free energy functional~\cite{RosenfeldMolPhys1998}. Mathematically their main result can be expressed as
\begin{equation}\label{RT1}
u_{\rm th}\simeq \alpha(\rho)\left(\frac{T}{T_{\rm fr}}\right)^{-2/5},
\end{equation}
where $u_{\rm th}$ is the thermal correction (i.e. the temperature dependent term) to the excess energy per particle in units of system temperature $T$, $T_{\rm fr}$ is the temperature at freezing, $\rho$ is the density, 
and $\alpha(\rho)$ is a density- and system-dependent parameter. Sometimes a single constant $\alpha\sim 3$ is appropriate~\cite{RosenfeldPRE2000,KhrapakPRE02_2015,KhrapakJCP2015}. 

The RT thermal energy scaling has important implications. For example, the RT scaling gives direct access to a very important thermodynamic quantity -- heat capacity~\cite{TrachenkoBook,BolmatovSciRep2012,KryuchkovPRL2020}. By virtue of the thermodynamic identity $c_{\rm v}^{\rm ex}=\partial (T u_{\rm th})/\partial T|_{\rho}$ we obtain for the reduced excess isochoric heat capacity 
\begin{equation}\label{RT2}
c_{\rm v}^{\rm ex}\simeq \frac{3}{5}\alpha(\rho) \left(\frac{T}{T_{\rm fr}}\right)^{-2/5}.
\end{equation}  
The accuracy of the RT scaling in the form of Eq.~(\ref{RT2}) has been examined for 18 model liquids in Ref.~\cite{IngebrigtsenJCP2013}. It has been observed that the RT expression is a better approximation for liquids with strong correlations between equilibrium fluctuations of virial and potential energy, known as “Roskilde-simple” liquids~\cite{IngebrigtsenPRX2012,DyreJPCB2014}. Later in Ref.~\cite{MausbachPRE2018} it has been demonstrated that the isomorph character of the freezing line and RT temperature scaling in the Lennard-Jones (LJ) fluid are consistent only in rather narrow parameter regime~\cite{MausbachPRE2018}.     

Furthermore, assuming that the scaling of Eq.~(\ref{RT1}) operates almost all the way to the high-temperature limit, Rosenfeld obtained the expression for the excess entropy~\cite{RosenfeldPRE2000}
\begin{equation}\label{RT3}
s_{\rm ex}\simeq s_{\rm ex}^{\rm fr}+\frac{3}{2}\alpha(\rho)\left[1-\left(\frac{T}{T_{\rm fr}}\right)^{-2/5}\right],
\end{equation} 
where $s_{\rm ex}^{\rm fr}$ is the excess entropy at the freezing point. Originally, Rosenfeld used $\alpha(\rho)=3$ and pointed out that $s_{\rm ex}^{\rm fr}\simeq -4$ for many soft interactions such as inverse power and Yukawa potentials~\cite{RosenfeldPRE2000}. Expression (\ref{RT3}) provides direct connection between the excess entropy and freezing temperature scaling of transport coefficients~\cite{RosenfeldPRE2000,RosenfeldJPCM2001} and thus represents an important result in the theory of the liquid state.

There is, nevertheless, some inconsistency in the relation between the excess entropy and temperature suggested by Eq.~(\ref{RT3}). Clearly it is accurate in the vicinity of the freezing transition, provided $s_{\rm ex}^{\rm fr}$ and $\alpha(\rho)$ are chosen appropriately. However, it becomes progressively inaccurate as the temperature increases, because it does not ensure a correct asymptote $s_{\rm ex}\rightarrow 0$ as $T\rightarrow \infty$ in the general case. One can fix this issue by enforcing $\alpha(\rho) =-(2/3)s_{\rm ex}^{\rm fr}$, so that
\begin{equation}\label{RTOld}
s_{\rm ex}\simeq s_{\rm ex}^{\rm fr}\left(\frac{T}{T_{\rm fr}}\right)^{-2/5}.
\end{equation}
However, there is no guarantee that this form (which we will refer to as the ``original'' RT scaling) would best describe the actual dependence of the excess entropy on temperature. 

The purpose of this paper is to clarify this point. Taking the Lennard-Jones (LJ) model fluid as a representative example, we calculate the excess entropy dependence on temperature along isochores. We consider densities above the triple point density so that $T_{\rm fr}(\rho)$ exists and the freezing temperature scaling can be tested. Using this calculation we document that the original RT scaling resulting in Eq.~(\ref{RT3}) does not represent the best option for describing the dependence of $s_{\rm ex}$ on $T$. A modified RT scaling, formulated in terms of the excess entropy itself, appears to provide a better description. In particular, we demonstrate that the scaling
\begin{equation}\label{RTNew}
s_{\rm ex}\simeq s_{\rm ex}^{\rm fr}\left(\frac{T}{T_{\rm fr}}\right)^{-1/3}
\end{equation}           
is much more consistent with the results of our calculation in the extended temperature range. We  study the specific heat of the LJ fluid and find that with respect to heat capacity a transition between modified and original RT scaling takes place as the density increases from the triple-point value. We further examine the specific heat of liquefied noble gases, such as neon, argon, krypton and xenon, as well as of a liquid metal represented by mercury. We find that both the original and modified scaling can operate without any obvious systematic indication with regard which of them should be employed and under which conditions. This leads us to propose a generalized RT scaling for the specific heat of dense simple fluids.

\section{Lennard-Jones fluid}

The Lennard-Jones (LJ) system is one of the most popular and extensively studied models in condensed matter, because it combines relative simplicity with adequate {\it qualitative} approximation of interactions in simple atomic substances, such as noble gases (including corresponding liquids and solids).  
The LJ potential is 
\begin{equation}
\phi(r)=4\epsilon\left[\left(\frac{\sigma}{r}\right)^{12}-\left(\frac{\sigma}{r}\right)^{6}\right], 
\end{equation}
where  $\epsilon$ and $\sigma$ are the energy and length scales (also referred to as conventional LJ units). The reduced density and temperature expressed in LJ units are $\rho^*=\rho\sigma^3$, $T^*=T/\epsilon$.  

The phase diagram, thermodynamic properties, and transport coefficients of the LJ fluids have been extensively investigated in the literature over many decades and are very well known. We remind the location of important reference points in the LJ system phase diagram in $(\rho^*, T^*)$ plane. The critical point density and temperature are $\rho^*\simeq 0.316$ and $T^*\simeq 1.326$~\cite{HeyesJCP2019}, respectively. The location of the gas-liquid-solid triple point is characterized by $\rho^*\simeq 0.846$ and $T^*\simeq 0.694$~\cite{SousaJCP2012}. We consider only densities above the triple point density, corresponding to the dense liquid and fluid regimes. In this regime the solid-like vibrational mechanism of atomic dynamics dominates on sufficiently short time scales~\cite{KhrapakPhysRep2024}.

The fluid-solid phase transition in the LJ fluid has also been extensively investigated, see Refs.~\cite{AgrawalMolPhys1995_2,HansenPRA1970,RosenfeldCPL1976,BarrosoJCP2002,KhrapakJCP2011_2} and references therein for some relevant examples. For our present purpose we take the temperatures and densities at the fluid side of the fluid-solid coexistence tabulated in Ref.~\cite{SousaJCP2012}. In particular, we consider seven isochores with densities $\rho^*\simeq 0.8626$, 0.9201, 1.0020, 1.0646, 1.1611, 1.2370, 1.2995. These correspond to the freezing density at temperatures $T^*=0.75$, 1.0, 1.5, 2.0, 3.0, 4.0, and 5.0, respectively~\cite{SousaJCP2012}.          


\section{Excess entropy on isochores}

We calculate the excess entropy of the LJ fluid with the help of the equation of state (EoS) developed by Thol {\it et al}.~\cite{Thol2016}. This EoS provides relative good accuracy and is convenient in practical implementation. It is based on simulations covering the homogeneous fluid region with $0.7 < T^* < 9$, $\rho^* < 1.08$. Extrapolation to the freezing line at $T^*<5$ is justified, as an analysis of the fluid-solid coexistence pressure indicates~\cite{KhrapakJPCL2022}. Extrapolation to even higher temperatures leads to relatively small deviations from Monte Carlo simulation results~\cite{MausbachPRE2018}.

\begin{figure}
\includegraphics[width=8.5cm]{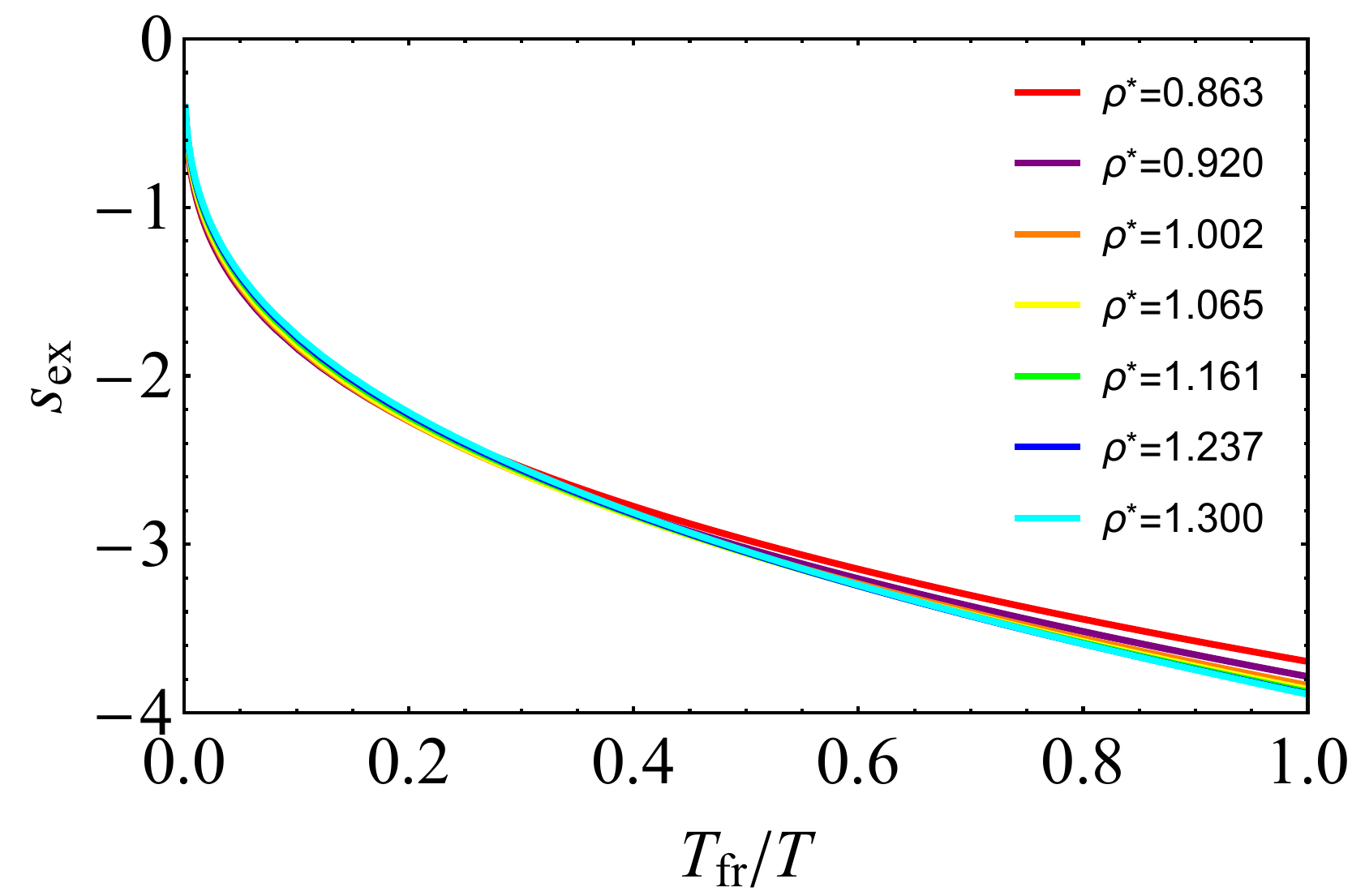}
\caption{(Color online) Reduced excess entropy $s_{\rm ex}$ as a function of reduced temperature $T_{\rm fr}/T$ along several LJ isochores (see the legend). The excess entropy is calculated from the EoS by Thol {\it et al.}~\cite{Thol2016}. Freezing line data are from Ref.~\cite{SousaJCP2012}.}
\label{FigEntropy}
\end{figure} 

The calculated excess entropy along the seven isochores specified above are plotted in Fig.~\ref{FigEntropy} as a function of the inverse reduced temperature $T_{\rm fr}/T$. We observe a rather good coincidence between different curves at temperatures exceeding twice the freezing temperature. On approaching the freezing line, slight deviations between different curves are observable. These deviations are not unexpected. Freezing and melting lines can be considered as having identical structure if properly scaled. This point of view was discussed already by Ross when generalizing Lindemann's melting law~\cite{RossPR1969}. Modern theories operate with isomorph lines on the phase diagram. Recent developments show that in many liquids and solids the so-called isomorphs exist, along which structure and dynamics in properly reduced units are invariant to a good approximation~\cite{DyreJPCB2014}. One of the conventional definitions of isomorphs is the constancy of the excess entropy along that line. Careful investigation shows that the freezing and melting lines in the LJ system are close to isomorphs, but are not exact isomorphs~\cite{PedersenNatCom2016}. Some variation in the thermodynamic and transport properties that are considered isomporph invariants along the LJ freezing line is observed~\cite{PedersenNatCom2016,CostigliolaPCCP2016}. This concerns the excess entropy as well. 

The dependence of $s_{\rm ex}^{\rm fr}$ on $T^*$ along the freezing line of the LJ system for $T^*<5$, based on the EoS by Thol {\it et al}.~\cite{Thol2016} and freezing line input from Ref.~\cite{SousaJCP2012}, is shown in Fig.~2 of Ref.~\cite{KhrapakJCP2022_1}. The excess entropy at freezing is $s_{\rm ex}^{\rm fr}\simeq -3.65$ at the triple point and saturates at $s_{\rm ex}^{\rm fr}\simeq -3.9$ as the temperature increases. Slightly higher values (saturation at $s_{\rm ex}^{\rm fr}\simeq -3.6$ in the high temperature limit) have been reported in Ref.~\cite{MausbachPRE2018}. The difference may originate from the deviation in the freezing line data used in these works. This difference is not essential for the present consideration, because $s_{\rm ex}^{\rm fr}$ merely serves as an adjustable parameter, as will become evident soon. 

\begin{figure}
\includegraphics[width=8.5cm]{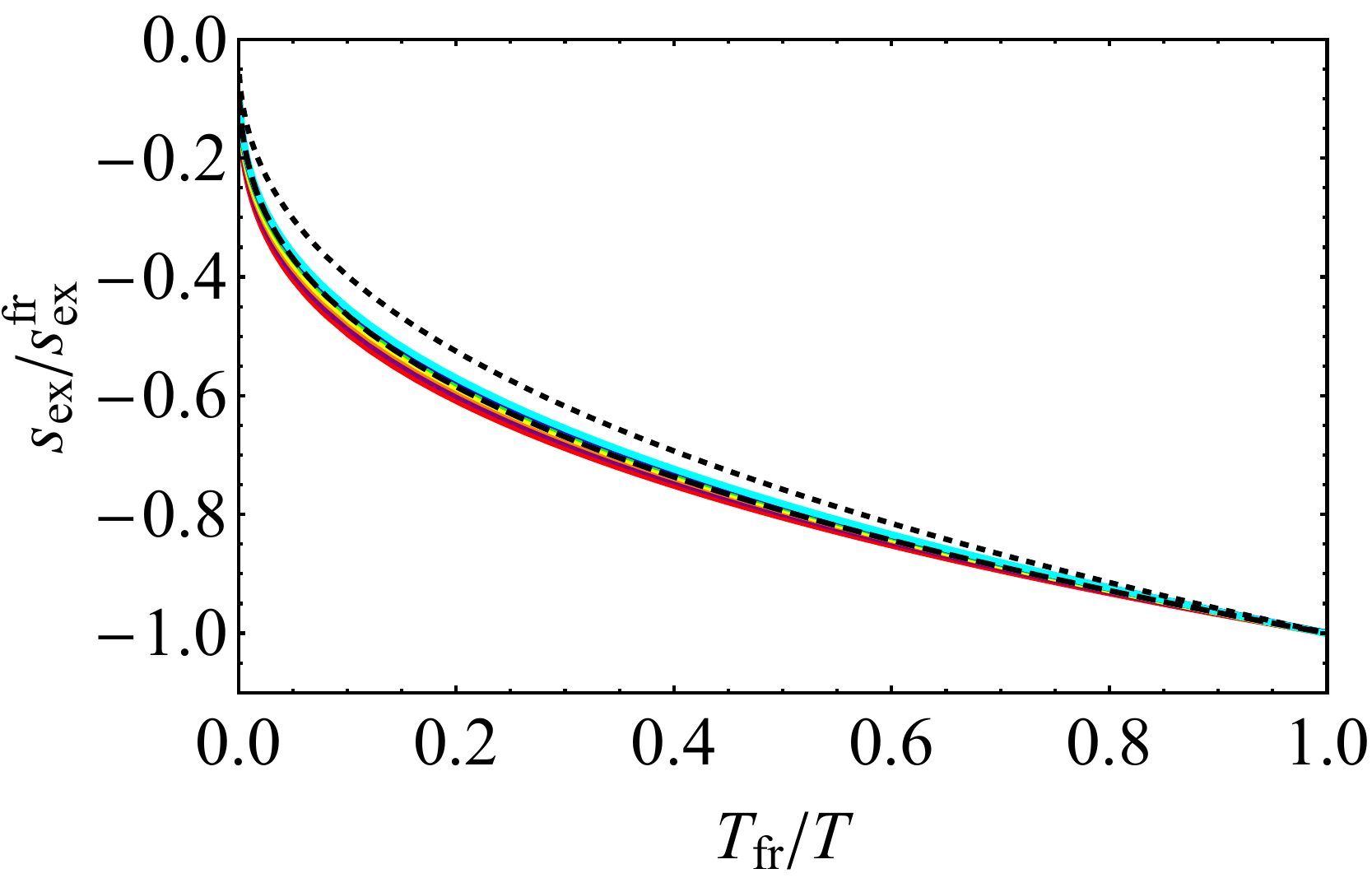}
\caption{(Color online) The ratio $s_{\rm ex}/s_{\rm ex}^{\rm fr}$ as a function of reduced temperature $T_{\rm fr}/T$ along the same isochores as in Fig.~\ref{FigEntropy}. The excess entropy is calculated from the EoS by Thol {\it et al.}~\cite{Thol2016}. The dashed line corresponds to the modified RT scaling of Eq.~(\ref{RTNew}). The dotted line corresponds to the original RT scaling of Eq.~(\ref{RTOld}). The color scheme is the same as in Fig.~\ref{FigEntropy}.}
\label{FigEntropyUni}
\end{figure} 

To eliminate the effect of variation of the excess entropy along the freezing line, we plot the dependence of the ratio $s_{\rm ex}/s_{\rm ex}^{\rm fr}$ in Fig.~\ref{FigEntropyUni}. We observe much better coincidence of curves corresponding to different isochores. We also see that the scaling of Eq.~(\ref{RTNew}), shown by the dashed curve,  describes the dependence better than that based on the original RT scaling, Eq.~(\ref{RTOld}), depicted by the dotted curve. The latter is acceptable in the vicinity of freezing, but the deviations are significant at higher temperatures. 

The proposed modification of the RT scaling has important implications, which are discussed in the next sections.

\section{Heat capacity of the LJ fluid}           

Heat capacity can be calculated from the thermodynamic identity $c_{\rm v}^{\rm ex}=T(\partial s_{\rm ex}/\partial T)_{\rho}$. Taking into account the modified RT scaling of Eq.~(\ref{RTNew}) and a rather weak dependence of $s_{\rm ex}^{\rm fr}$ on temperature, it is tempting to assume that a scaling $c_{\rm v}^{\rm ex}\propto (T/T_{\rm fr})^{-1/3}$ should hold. It is, however, not so straightforward, as we demonstrate below. 
 
\begin{figure}
\includegraphics[width=8.5cm]{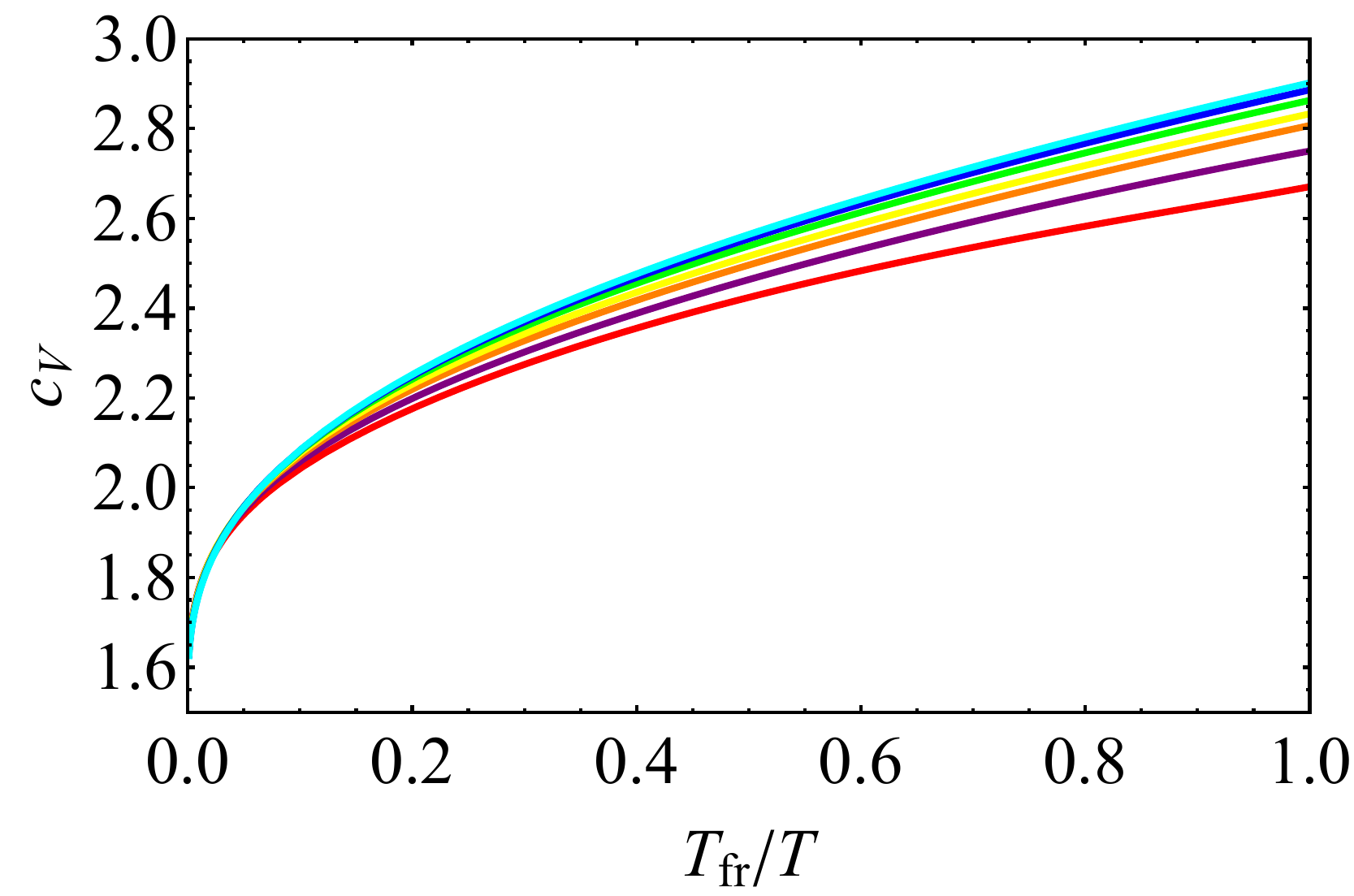}
\caption{(Color online) Reduced isochoric specific heat $c_{\rm v}$ as a function of reduced temperature $T_{\rm fr}/T$ along the same LJ isochores as in Fig.~\ref{FigEntropy}. The color scheme is identical to that in Fig.~\ref{FigEntropy}. The specific heat is calculated using the EoS by Thol {\it et al.}~\cite{Thol2016}. Freezing line data are from Ref.~\cite{SousaJCP2012}.} 
\label{FigCv}
\end{figure}

In Fig.~\ref{FigCv} the dependence of the specific heat at constant volume on the ratio  $T_{\rm fr}/T$ is plotted, as evaluated from the EoS by Thol {\it et al.}~\cite{Thol2016}. On approaching to the freezing transition the magnitude of $c_{\rm v}$ approaches the value of $3.0$ as could be expected from Dulong-Petit law. In the high-temperature limit $c_{\rm v}$ approaches the ideal gas value of $3/2$. The curves are to some extent quasi-universal, but a notable increase of $c_{\rm v}$ with density at the freezing point is clearly observed. 

\begin{figure}
\includegraphics[width=8.5cm]{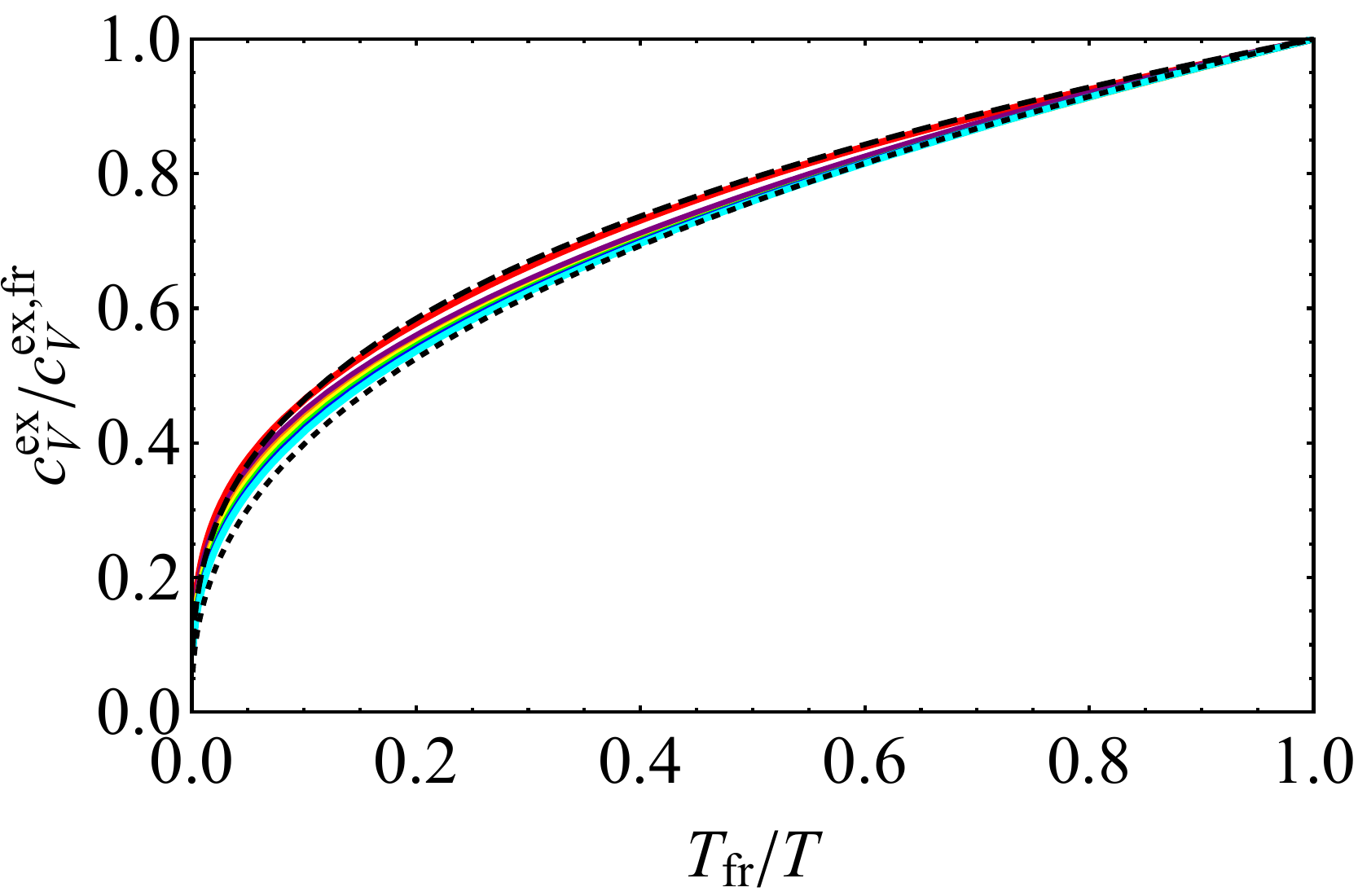}
\caption{(Color online) Ratio of the excess specific heat at constant volume, divided by its value at the freezing point, $c_{\rm v}^{\rm ex}/c_{\rm v}^{\rm ex, fr}$, as a function of the reduced temperature $T_{\rm fr}/T$. Results are presented for the same isochores as in Fig.~\ref{FigCv} and the color scheme is identical. The excess specific heat is calculated from the EoS by Thol {\it et al.}~\cite{Thol2016}. Freezing line data are from Ref.~\cite{SousaJCP2012}. The dashed curve corresponds to the modified RT scaling $c_{\rm v}^{\rm ex}\propto (T/T_{\rm fr})^{-1/3}$. The dotted curve shows the  original RT scaling $c_{\rm v}^{\rm ex}\propto (T/T_{\rm fr})^{-2/5}$.}
\label{FigCvUni}
\end{figure}

To eliminate the variation of $c_{\rm v}$ with density in the vicinity of freezing, we plot the excess contribution to specific heat, divided by its value at the freezing point, in Fig.~\ref{FigCvUni}. The coincidence between the curves is now much better pronounced. We also plot the outcome of the modified RT scaling $c_{\rm v}^{\rm ex}\propto (T/T_{\rm fr})^{-1/3}$ by the dashed curve and that of the original RT scaling $c_{\rm v}^{\rm ex}\propto (T/T_{\rm fr})^{-2/5}$ by the doted curve. It appears that the modified RT scaling describe better the data for the near-triple-point density, while the original RT scaling is closer to the data corresponding to higher densities and temperatures. The curves for intermediate densities and temperatures lie just in between the modified and original RT scaling.       

In the next section we well see how these tendencies correlate with the actual behaviour of specific heat in real liquids. 

\section{Real fluids}

While the heat capacity of gases and solids is well understood and discussed in numerous textbooks on statistical physics and physics of the solid state, no convincing theory of the heat capacity of liquids exists~\cite{TrachenkoBook}. Not surprisingly, this remains a topic of considerable current interest and debate~\cite{BolmatovSciRep2012,ProctorPoF2020,KryuchkovPRL2020,
BaggioliPRE2021,SchirmacherPRE2022}. Therefore, the question whether the original RT scaling can be helpful for simple practical estimates or modifications are required represents considerable interest. 

We have examined the recommended data for the isochoric specific heat of four liquefied noble gases -- neon, argon, krypton, and xenon. These data are provided in the Institute of Standards and Technology (NIST) Reference Fluid Thermodynamic and Transport Properties Database (REFPROP 10)\cite{Refprop}. For each fluid, two to three isochores, one of which emanates from the triple point are considered. 
Helium liquid is not considered, because in this case quantum effects are expected to play a significant role~\cite{HansenPR1969}. The individual models for the EoS for each species implemented
in REFPROP 10 can be found in project documentation~\cite{Refprop} and in Refs.~\cite{Thol2018,Tegeler1999,Lemmon2006}.

\begin{figure}
\includegraphics[width=8.5cm]{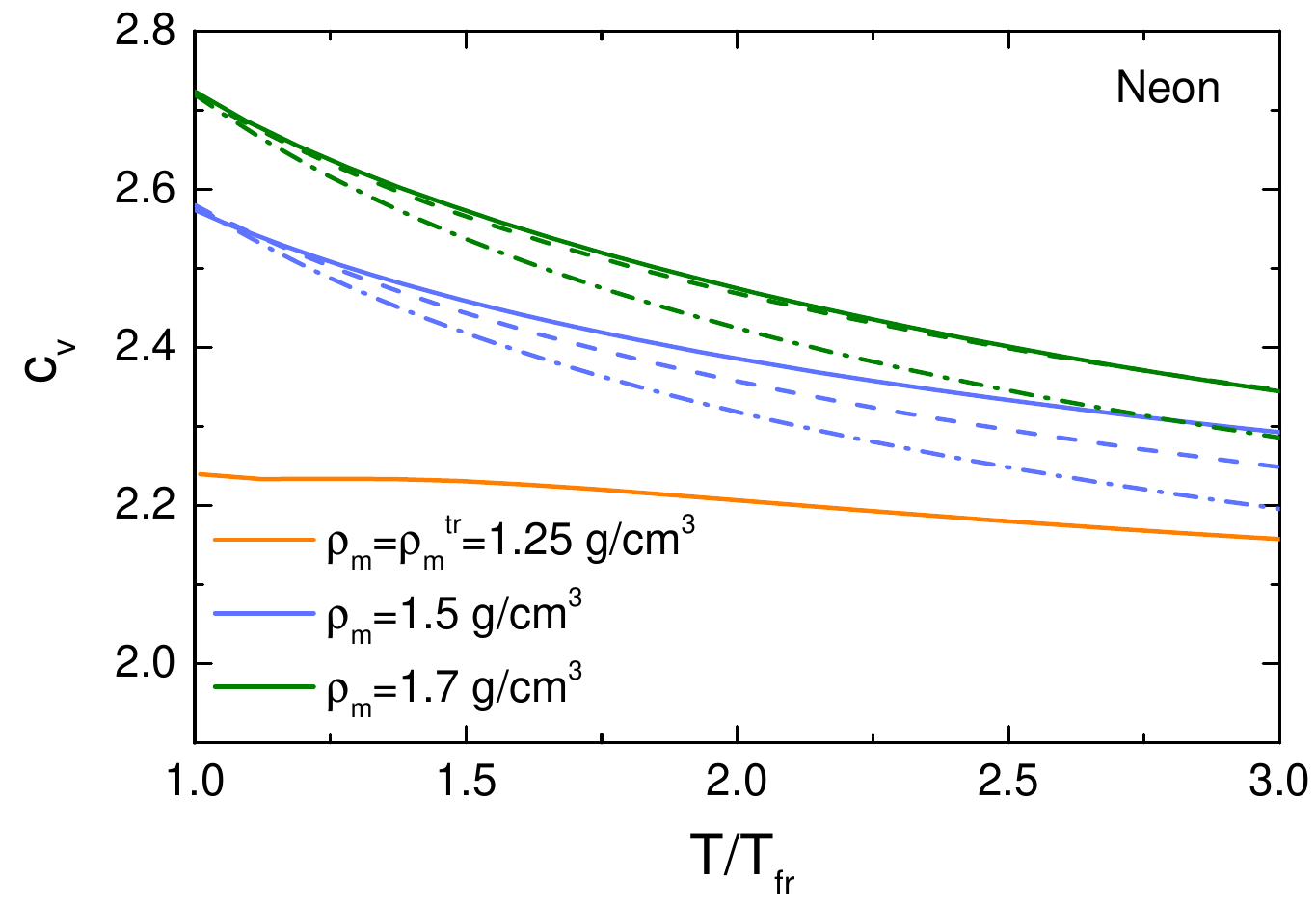}
\caption{(Color online) Reduced specific heat at constant volume $c_{\rm v}$ of liquefied neon as a function of reduced temperature $T/T_{\rm fr}$ along three isochores $\rho_m=1.25$, 1.5, and 1.7 g/cm$^3$. Solid curves are the recommended values from the NIST database~\cite{Refprop}. The dashed curves correspond to the modified RT scaling $c_{\rm v}^{\rm ex}\propto (T_{\rm fr}/T)^{1/3}$. The dash-dotted curves denote the original RT scaling $c_{\rm v}^{\rm ex}\propto (T_{\rm fr}/T)^{2/5}$.}
\label{FigNeon}
\end{figure} 

Reduced specific heat $c_{\rm v}$ of neon liquid along the isochores $\rho_{\rm m} = 1.25$, 1.5 amd 1.7 g/cm$^3$ are shown in Fig.~\ref{FigNeon}. The behaviour along the triple-point isochore is unconventional, no expected freezing temperature scaling is present. This might be attributed to quantum effects, since the average interatomic separation is comparable to  the thermal de Broglie wavelength in neon near the triple point temperature~\cite{KhrapakPoF2023}. As the density and hence temperature increases, the role of quantum effects diminishes and the expected dependence of $c_{\rm v}$ 
on $T$ is realized. The modified RT scaling $c_{\rm v}^{\rm ex}\propto (T_{\rm fr}/T)^{1/3}$ is plotted by the dashed curves. The original RT scaling $c_{\rm v}^{\rm ex}\propto (T_{\rm fr}/T)^{2/5}$ is shown by the dash-dotted curves. We observe that modified RT scaling is in considerably better agreement with the recommended data than the original one. This is particularly true for the highest density considered.     

\begin{figure}
\includegraphics[width=8.5cm]{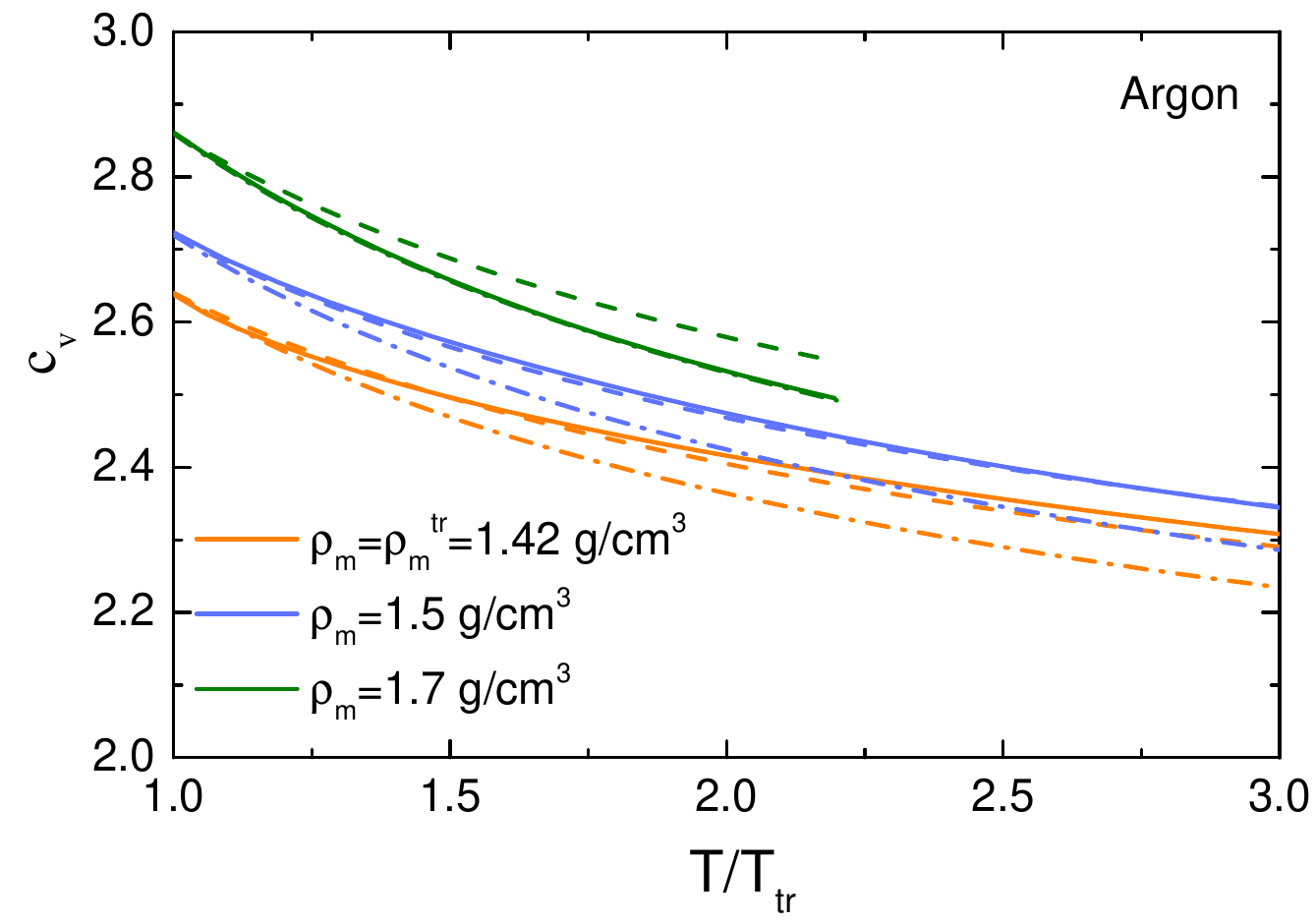}
\caption{(Color online) Reduced specific heat at constant volume $c_{\rm v}$ of liquefied argon as a function of reduced temperature $T/T_{\rm fr}$ along three isochores $\rho_m=1.42$, 1.5, and 1.7 g/cm$^3$. Solid curves are the recommended values from the NIST database~\cite{Refprop}. The dashed curves correspond to the modified RT scaling $c_{\rm v}^{\rm ex}\propto (T_{\rm fr}/T)^{1/3}$. The dash-dotted curves denote the original RT scaling $c_{\rm v}^{\rm ex}\propto (T_{\rm fr}/T)^{2/5}$.}
\label{FigArgon}
\end{figure} 

Figure~\ref{FigArgon} demonstrated the dependence of the reduced specific heat on reduced temperature in argon along isochores  $\rho_{\rm m} = 1.42$, 1.5 amd 1.7 g/cm$^3$. The behaviour is conventional along all three isochores. For the two lower density isochores, the modified RT scaling is in closer agreement with the reference data. However, for the higher density isochore, the reference data are perfectly described by the original RT scaling.

\begin{figure}
\includegraphics[width=8.5cm]{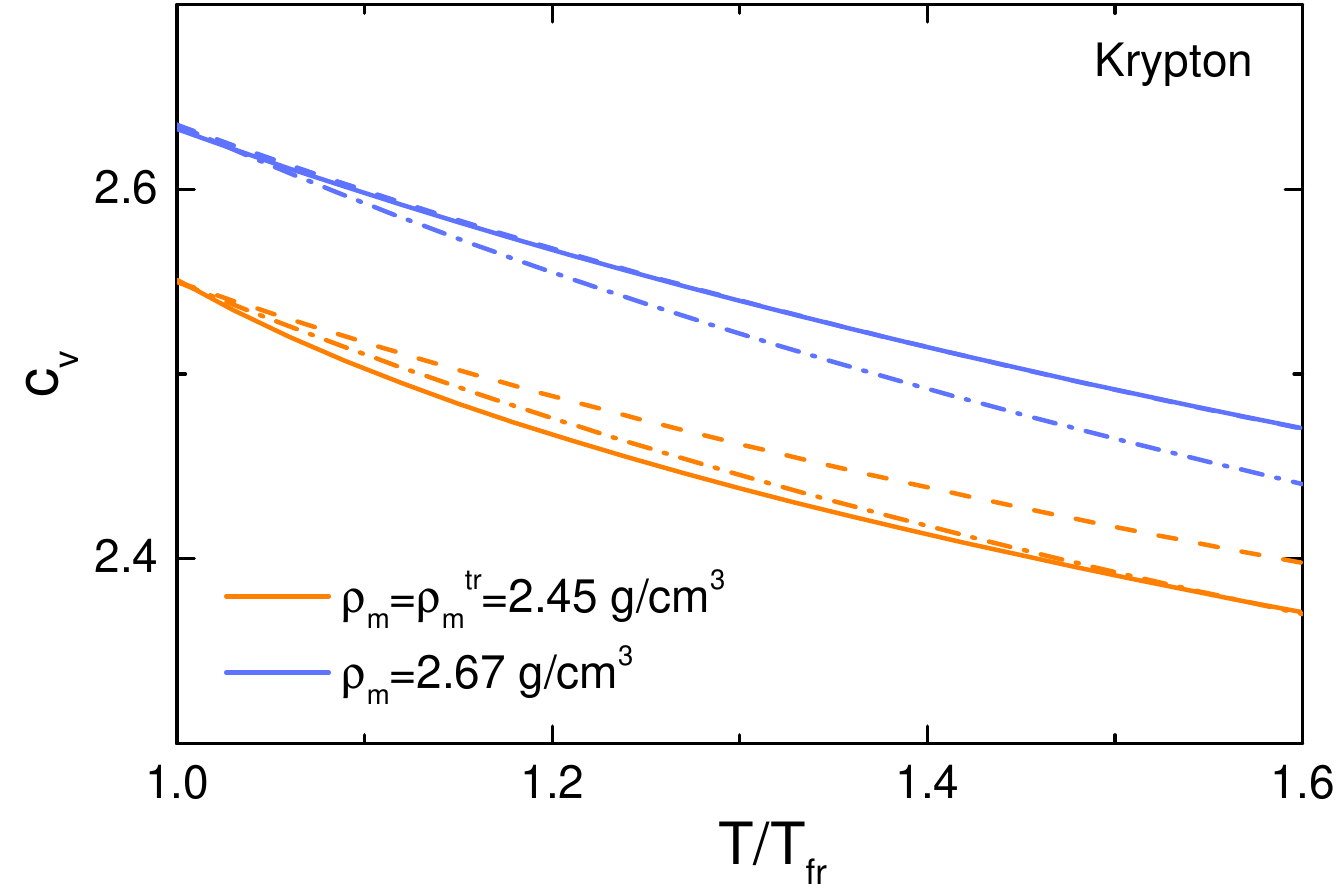}
\caption{(Color online) Reduced specific heat at constant volume $c_{\rm v}$ of liquefied krypton as a function of reduced temperature $T/T_{\rm fr}$ along two isochores $\rho_m=2.45$, and 2.67 g/cm$^3$. Solid curves are the recommended values from the NIST database~\cite{Refprop}. The dashed curves correspond to the modified RT scaling $c_{\rm v}^{\rm ex}\propto (T_{\rm fr}/T)^{1/3}$. The dash-dotted curves denote the original RT scaling $c_{\rm v}^{\rm ex}\propto (T_{\rm fr}/T)^{2/5}$.}
\label{FigKrypton}
\end{figure} 

For krypton, specific heat data for two isochores, $\rho_{\rm m} = 2.45$ and 2.67 g/cm$^3$ are shown in Fig.~\ref{FigKrypton}. For the lower density isochore, the original RT scaling is closer to the reference data. On the higher density isochore the modified RT scaling almost ideally coincides with the recommended values.

\begin{figure}
\includegraphics[width=8.5cm]{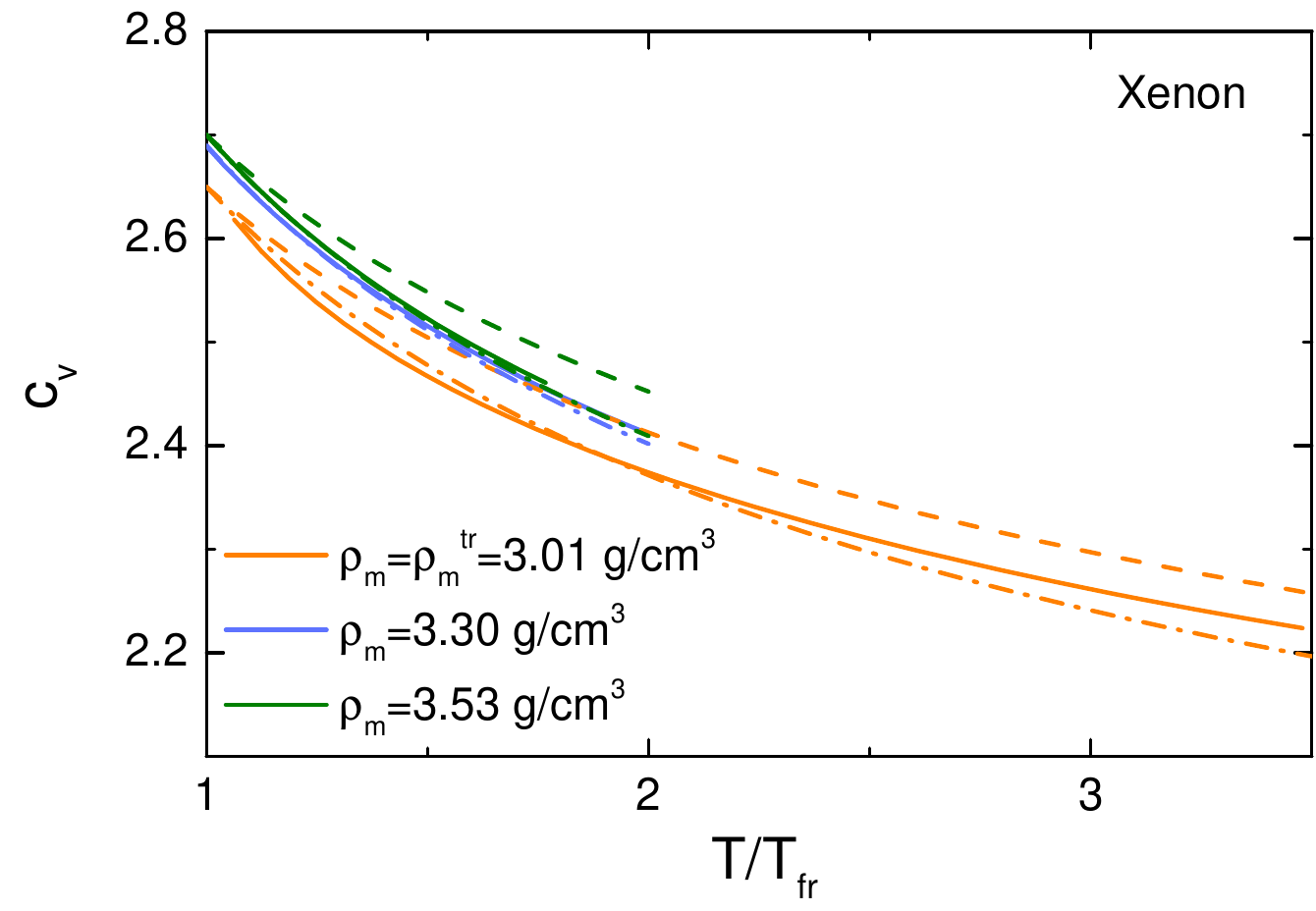}
\caption{(Color online) Reduced specific heat at constant volume $c_{\rm v}$ of liquefied xenon as a function of reduced temperature $T/T_{\rm fr}$ along along three isochores $\rho_m=3.01$, 3.30, and 3.53 g/cm$^3$. Solid curves are the recommended values from the NIST database~\cite{Refprop}. The dashed curves correspond to the modified RT scaling $c_{\rm v}^{\rm ex}\propto (T_{\rm fr}/T)^{1/3}$. The dash-dotted curves denote the original RT scaling $c_{\rm v}^{\rm ex}\propto (T_{\rm fr}/T)^{2/5}$.}
\label{FigXenon}
\end{figure} 

Figure~\ref{FigXenon} shows the dependence of $c_{\rm v}$ on $T/T_{\rm fr}$ in xenon along three isochores  $\rho_{\rm m} = 3.01$, 3.30 amd 3.53 g/cm$^3$. In this case the reference data along all three isochores can be somewhat better described by the original RT scaling $\propto (T/T_{\rm fr})^{-2/5}$.  

\begin{figure}
\includegraphics[width=8.5cm]{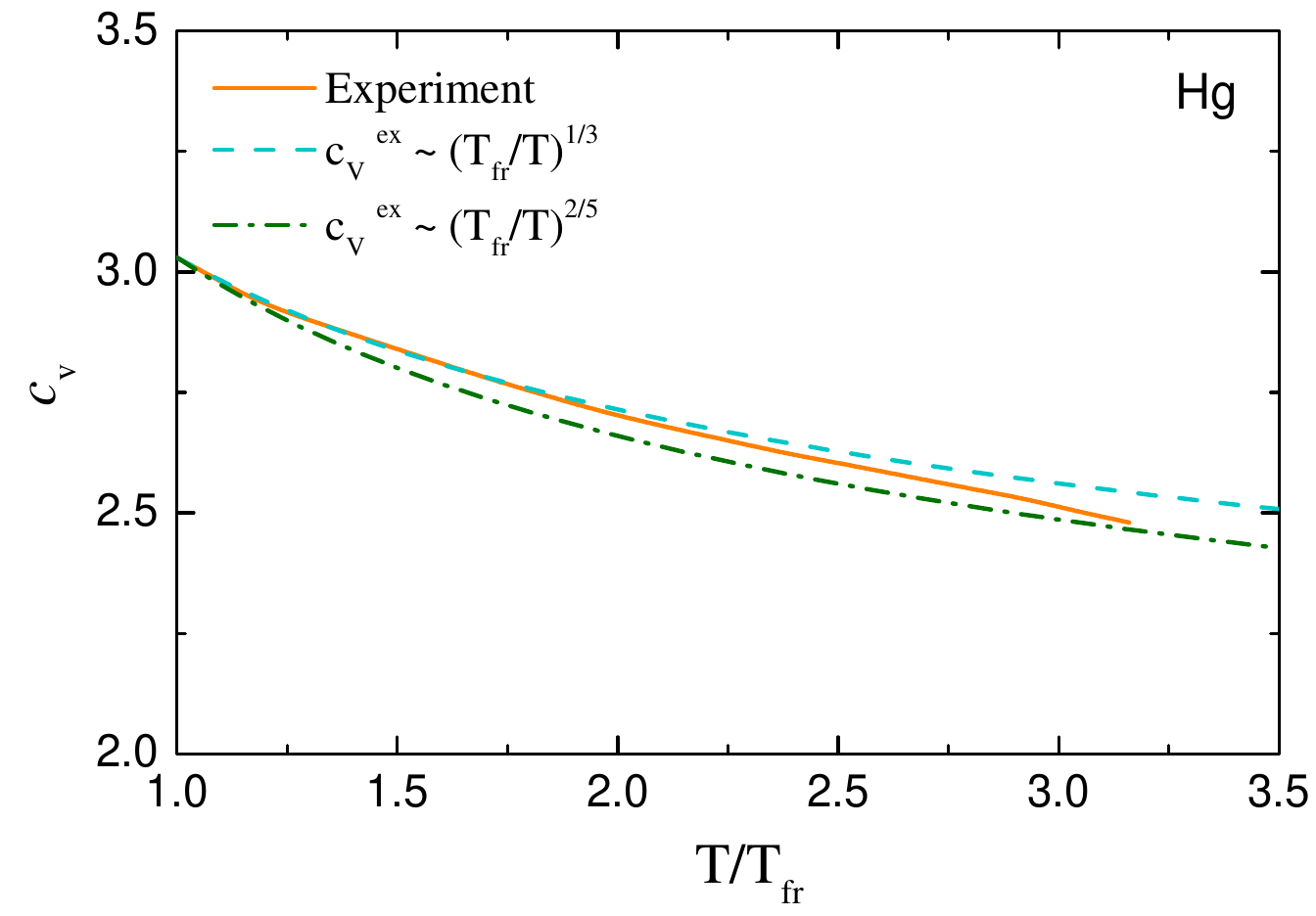}
\caption{(Color online) Reduced isochoric specific heat $c_{\rm v}$ of liquid mercury (with the electronic contribution subtracted) as a function of reduced temperature $T/T_{\rm fr}$ along an isochore $\rho_m=13.69$ g/cm$^3$. The experimental data shown by the solid line correspond to those appearing in Fig.~3 of Ref.~\cite{WallacePRE1998}. The dashed and dash-dotted curves correspond to the modified and original RT scaling, respectively.}
\label{FigHg}
\end{figure}

In addition to liquefied noble gases, we show the specific heat $c_{\rm v}$ (with electronic contribution subtracted) in mercury (Hg) as a function of $T/T_{\rm fr}$ in Fig.~\ref{FigHg}. Experimental data correspond to those presented in Fig.~3 of Ref.~\cite{WallacePRE1998}. The dashed and dash-dotted curves are calculated using the modified and original RT scaling, respectively. We see that near the fluid-solid phase transition ($T\lesssim 2T_{\rm fr}$) the modified RT scaling is more appropriate. For higher temperatures the experimental results gradually tend to the original RT asymptote $\propto (T/T_{\rm fr})^{-2/5}$.   

\section{Discussion and conclusion}

Our first important observation is that the excess entropy of the LJ fluids exhibits a universal scaling of the form (\ref{RTNew}) for densities exceeding the triple point density. This can be considered as a modified RT scaling. Modified RT scaling provides easy access to the transport properties of dense LJ fluids. For instance, expressions proposed in Ref.~\cite{BellJPCB2019} can be used to relate the coefficients of self-diffusion, shear viscosity and thermal conductivity to the magnitude of the excess entropy. 

Based on the functional form of the modified RT scaling and rather weak variation of the excess entropy along the freezing line, one can expect that the excess contribution to the isochoric specific heat also scales as $\propto (T_{\rm fr}/T)^{1/3}$. Careful examination demonstrates however, that for the LJ fluid, this modified RT scaling is more appropriate at near critical point densities. For higher densities the original RT scaling  $\propto (T_{\rm fr}/T)^{2/5}$ is approached (see Fig.\ref{FigCvUni}). 

Another useful observation is that the reduced specific heat of the LJ fluid at freezing increases with density towards the value $3.0$, expected from the Dulong-Petit law. This tendency is fully reproduced in liquefied noble gases such as neon, argon, krypton and xenon. The transition from the modified RT scaling to the original RT scaling upon increasing density observed in the LJ fluid is not fully supported by the data from the NIST database. The picture arising from the LJ model is well reproduced only in the case of argon. For neon the modified RT scaling is usually more accurate. For krypton the modified RT scaling shows better accuracy at high density, while the near-triple-point density data are better described by the original RT scaling. The data for xenon are better consistent with the original RT scaling. Finally, the experimentally measured dependence of $c_{\rm v}$ on $T/T_{\rm fr}$ in liquid mercury is consistent with the modified RT scaling in the vicinity of freezing, but tends to the original RT scaling at higher temperatures. Thus, no regular internally consistent picture (within the present context) emerges from the analysis of heat capacity data in real liquids.   

Taking these observations into account we propose to describe the specific heat of dense fluids using the expression
\begin{equation}\label{GeneralizedRT}
c_{\rm v}^{\rm ex}\propto \left(\frac{T_{\rm fr}}{T}\right)^{\alpha},
\end{equation}               
where $\alpha$ is a density- and material-dependent adjustable parameter. The actual values of $\alpha$ for the isochores in neon, argon, krypton and xenon considered above, obtained by fitting the NIST data up to $T/T_{\rm fr}\lesssim 2.0$ in neon, argon and xenon, and up to $T/T_{\rm fr}\lesssim 1.6$ in krypton are summarized in Table~\ref{Tab1}. From the presented evidence we might conclude that the parameter $\alpha$ belongs to a relatively narrow window between $\simeq 1/4$ and $\simeq 2/5$. There are, however, strong indications that in the case of soft repulsive Yukawa (screened Coulomb) potential, playing important role in the context of plasma-related and colloidal systems, the exponent $\alpha=1/2$ is more appropriate~\cite{KhrapakPRE09_2024}. This might indicate that the exponent $\alpha$ increases with the steepness of the interaction potential. However, the role of attractive component in the interaction potential also deserves attention. In the special case of hard-sphere fluids, neither the excess specific heat nor the excess entropy depend on the temperature. Formally, this corresponds to $T_{\rm fr}=0$ in Eq.~(\ref{GeneralizedRT}). The value of the exponent $\alpha$ is irrelevant in this case. 

\begin{table}
\caption{\label{Tab1} Exponents $\alpha$ of the generalized RT scaling (\ref{GeneralizedRT}) as obtained by fitting the specific heat $c_{\rm v}$ data from the NIST database~\cite{Refprop} for neon, argon, krypton, and xenon liquids along the isochores considered above.}
\begin{ruledtabular}
\begin{tabular}{lcc}
Liquid & Mass density $\rho_m$ (g/cm$^3$) & $\alpha$  \\ \hline
neon &  $1.50$  & $0.27$  \\  
neon & $1.70$ & $0.32$  \\
argon & $1.42$ & $0.33$ \\
argon & $1.50$ & $0.32$  \\
argon & $1.70$ & $0.40$ \\
krypton & $2.45$ & $0.42$ \\
krypton & $2.67$ & $0.33$ \\
xenon & $3.01$ & $0.43$ \\
xenon & $3.30$ & $0.38$ \\
xenon & $3.53$ & $0.39$ \\ 
\end{tabular}
\end{ruledtabular}
\end{table}   

Simple arguments, explaining why Eq.~(\ref{GeneralizedRT}) may work in moderately dense fluids and dense gases as well as how the exponent $\alpha$ can be related to the properties of the pairwise interaction potential have been discussed by Stishov~\cite{Stishov1980}. While these arguments are likely insufficient to justify its application to dense fluids near the fluid-solid phase transition, it probably makes sense to address this issue with the contemporary evidence on model and real liquids at hand.  

To conclude, an empirical relation (\ref{GeneralizedRT}), which can be considered as a {\it generalized RT scaling} for the specific heat of simple fluids is a major result of this work. The question concerning which material properties affect the parameter $\alpha$ and whether it can be predicted from general physical arguments requires further investigation. This would provide deep insight into physical properties of liquids and fluids across their phase diagrams.     

The authors have no conflicts of interest to disclose.

The data that support the findings of this study are available from the corresponding author, SK, upon reasonable request. 




\bibliography{SE_Ref}

\end{document}